\documentclass[prd,twocolumn,showpacs,showkeys,floatfix]{revtex4}
\usepackage{amsmath,amssymb}
\usepackage{latexsym}
\usepackage{bm}

\usepackage{hyperref}
\usepackage[ansinew]{inputenc}
\makeatletter
\def\eqnarray{\stepcounter{equation}\let\@currentlabel=\theequation
\global\@eqnswtrue
\global\@eqcnt\z@\tabskip\@centering\let\\=\@eqncr
$$\halign to \displaywidth\bgroup\@eqnsel\hskip\@centering
  $\displaystyle\tabskip\z@{##}$&\global\@eqcnt\@ne
  \hfil${\;##\;}$\hfil
  &\global\@eqcnt\tw@ $\displaystyle\tabskip\z@{##}$\hfil
   \tabskip\@centering&\llap{##}\tabskip\z@\cr}
\makeatother

\begin{document}
\title{LEMAITRE--TOLMAN--BONDI  DUST SPACETIMES:  SYMMETRY PROPERTIES AND SOME EXTENSIONS TO THE DISSIPATIVE CASE}
\author{L. Herrera}
\email{laherrera@cantv.net.ve}
\affiliation{Escuela de F\'{\i}sica, Facultad de Ciencias,
Universidad Central de Venezuela, Caracas, Venezuela.}
\author{A. Di Prisco}
\email{adiprisc@fisica.ciens.ucv.ve}
\affiliation{Escuela de F\'{\i}sica, Facultad de Ciencias,
Universidad Central de Venezuela, Caracas, Venezuela.}
\author{J. Ospino}\affiliation{Departamento de Matem\'atica Aplicada,  Universidad de Salamanca, Salamanca, Spain.}
\email{jhozcrae@usal.es}
\author{J. Carot}
\affiliation{Departament de  F\'{\i}sica, Universitat Illes Balears, E-07122 Palma de Mallorca, Spain}
\email{jcarot@uib.cat}
\begin{abstract}
We consider  extensions of Lemaitre--Tolman--Bondi (LTB) spacetimes to the dissipative case. For doing that we  previously carry out a systematic study on LTB. This study is based on two different  aspects of LTB. On the one hand, a symmetry property of  LTB will be presented. On the other  hand,  the description  of LTB in terms of some fundamental scalar functions (structure scalars)  appearing in the orthogonal splitting of Riemann tensor will be provided.  We shall consider as ``natural'' generalizations of LTB (hereafter referred to as GLTB) either those metrics  admitting some similar kind of symmetry as LTB, or those sharing  structure scalars with similar dependence on  the metric.

\end{abstract}

\date{\today}
\maketitle
\section{INTRODUCTION}
LTB dust models \cite{1, 2, 3} are among the oldest and most interesting solutions to Einstein equations. They describe spherically symmetric distribution of   inhomogeneous non--dissipative dust (see \cite{4, 5} for a detailed description of these spacetimes).

They have been used as cosmological models (see \cite{ns, sn, 8, 7, 6} and references therein), in the study of gravitational collapse and the problem of the cosmic censorship \cite{9, 10, 11, 12, J, m1, m2}, and in quantum gravity \cite{13, 14}.

A renewed interest in LTB has appeared, in  relation with  recent observations of type Ia supernovae, indicating that the  expansion of the universe is accelerating. Indeed, even if it is true that there is  general  consensus to invoke dark energy as a source of anti-gravity for understanding
the cosmic acceleration, it is also true that a growing number of researchers consider that inhomogeneities can account for the observed cosmic acceleration, without invoking dark energy (see \cite{Coley1, Coley2, 17p, 15, 7'', cel, 7N} and references therein).

Now, in spite of all their interest, LTB spacetimes present an important limitation, namely: they do not admit dissipative fluxes.
This is a serious shortcoming since   it is already an established fact that gravitational collapse is a
highly dissipative process (see \cite{Hs, Hetal, Mitra} and references
therein). This dissipation is required to account for the very large
(negative) binding energy of the resulting compact object (of the order
of $-10^{53} erg$).

 Indeed, it appears that the only plausible
mechanism to carry away the bulk of the binding energy of the collapsing
star, leading to a neutron star or black hole is neutrino emission
\cite{K}.

 Dissipation processes are usually treated invoking  two possible (opposite) approximations: diffusion and streaming out.

 In the diffusion approximation, it is assumed that the energy flux of
radiation (as that of
thermal conduction) is proportional to the gradient of temperature. In this regime, an equation of transport should be assumed in order to obtain the temperature distribution for each model. 

The diffusion approximation  is in general very sensible, since it applies whenever the mean free path of
particles responsible for the propagation of energy  is  very small as compared with the typical
length of the object, a circumstance found very often in astrophysical scenarios.

In fact, for a main sequence star such as the sun, the mean free path of
photons at the centre, is of the order of $2\, cm$. Also, the
mean free path of trapped neutrinos in compact cores of densities
about $10^{12} \, g.cm.^{-3}$ becomes smaller than the size of the stellar
core \cite{3n, 4n}.

Furthermore, the observational data collected from supernovae 1987A
indicate that the regime of radiation transport prevailing during the
emission process, is closer to the diffusion approximation than to the
streaming out limit \cite{5n}.

However in many other circumstances, the mean free path of particles transporting energy may be large enough so as to justify the  free streaming approximation. Therefore we
shall include simultaneously both limiting  cases of radiative transport (diffusion and streaming out), allowing us to describe a wide range of situations.

On the other hand,  at cosmological scales,  even if it is true that cold dark matter is non--collisional and strongly dominated by rest--mass, so that pressure
and heat flux terms (of kinetic nature) are negligible, it could be interesting to test the stability of conclusions and results based on the assumptions above, with respect to small deviations from those assumptions. In this sense, GLTB's with arbitrarily small  (but non--vanishing) dissipative fluxes, could be very helpful.

From all of  the above, the motivations to generalize LTB spacetimes as to admit dissipative fluxes are clearly justified.

Therefore, it is our goal in this manuscript to consider  possible generalizations of LTB spacetimes to the dissipative case. As it should be obvious, such  generalizations are not unique. As a ``qualitative'' guide in our endeavour we shall  look for  GLTB's ``as  similar'' as possible to LTB's. More precise definitions of what we mean by ``qualitative'' guide and ``as similar as possible'',  will be provided  later.

Thus for example,  exact solutions to Einstein equations describing dissipative geodesic fluids have been found in  \cite{19, 20, 22, 18, 17}, however all of them are  shear--free and we know that a distinct property of LTB is  it shear, accordingly we shall search for shearing GLTB. Shearing dissipative geodesic fluids may be found in \cite{21} and  \cite{24}, though they do not  become  LTB  in the non--dissipative case.

Our search will be based  on  two different types of arguments. On the one hand, symmetry arguments. We shall find a symmetry property of LTB spacetimes and we shall assume that the corresponding generalizations (GLTB) describing dissipative dust, share the same kind of symmetry. On the other hand, we shall describe LTB in terms of some scalar functions which emerge from the orthogonal splitting of the Riemann tensor. We shall assume that two of these scalar functions share the same form (with respect to metric functions) in both, LTB and GLTB.

In the specific case of localized  configurations we have to assume that our fluid distribution is bounded by a spherical surface. In order to avoid thin shells on such a boundary surface Darmois \cite{26} conditions should be imposed.

We would like to emphasize that, even though some specific examples are exhibited, our main goal in this work consists in providing different techniques  to obtain exact solutions representing geodesic radiating fluids whose properties are in some respect similar to LTB.

\section{FLUID DISTRIBUTION, KINEMATICAL VARIABLES AND BASIC EQUATIONS}
We consider a spherically symmetric distribution  of
geodesic fluid, which may be bounded by a spherical surface $\Sigma$, or not. The fluid is
assumed to be pure dust undergoing dissipation in the
form of heat flow (diffusion approximation) and outgoing null fluid (streaming out limit).

 Choosing comoving coordinates the general
 metric can be written (in the case of bounded configurations such a line element applies  to the fluid inside $\Sigma$) as

\begin{equation}
ds^2=-dt^2+B^2dr^2+R^2(d\theta^2+\sin^2\theta d\phi^2),
\label{1}
\end{equation}
where $B$ and $R$ are functions of $t$ and $r$ and are assumed
positive. We number the coordinates $x^0=t$, $x^1=r$, $x^2=\theta$
and $x^3=\phi$. Observe that  $B$ is dimensionless, whereas $R$ has the same dimension as $r$. Also observe that $t$ is the proper time.

The energy-momentum $T_{\alpha\beta}$ (inside $\Sigma$ if the system is bounded)
is assumed to have the form
\begin{equation}
T_{\alpha\beta}=\mu V_{\alpha}V_{\beta}+q_{\alpha}V_{\beta}+V_{\alpha}q_{\beta}+
\epsilon l_{\alpha}l_{\beta}, \label{3}
\end{equation}
where $\mu$ is the energy density, $q^{\alpha}$ the heat flux, $\epsilon$ the radiation density,
$V^{\alpha}$ the four-velocity of the fluid and $l^{\alpha}$ a  null four-vector.
These quantities satisfy

\begin{eqnarray}
V^{\alpha}V_{\alpha}=-1, \;\; V^{\alpha}q_{\alpha}=0, \;\;
 \nonumber \\ \;\;
l^{\alpha}V_{\alpha}=-1, \;\; l^{\alpha}l_{\alpha}=0.
\label{4}
\end{eqnarray}

Since we have chosen a comoving coordinate system, we have
\begin{eqnarray}
V^{\alpha}=A^{-1}\delta_0^{\alpha}, \;\;
q^{\alpha}=qB^{-1}\delta^{\alpha}_1, \;\;\nonumber \\
l^{\alpha}=A^{-1}\delta^{\alpha}_0+B^{-1}\delta^{\alpha}_1, \;\;
\label{5}
\end{eqnarray}
where  $q$ is a function of  $t$ and $r$,    $q^\alpha = q
\chi^\alpha$ and
$\chi^{\alpha}$ a unit four-vector along the radial direction, satisfying
\begin{equation}
\chi^{\alpha}\chi_{\alpha}=1,\;\;
\chi^{\alpha}V_{\alpha}=0, \;\;
\chi^{\alpha}=B^{-1}\delta^{\alpha}_1.
\label{chi}
\end{equation}

It may be more convenient to write (\ref{3}) in the form
\begin{equation}
T_{\alpha \beta} = \tilde{\mu} V_\alpha V_\beta + \tilde{q} \left(V_\alpha
\chi_\beta + \chi_\alpha V_\beta\right) +\epsilon \chi_\alpha \chi_\beta, \label{Tab}
\end{equation}
with
$$\tilde \mu= \mu+\epsilon,$$
$$\tilde q= q+\epsilon.$$

\subsection{Einstein equations}
For (\ref{1}) and (\ref{Tab}), Einstein equations

\begin{equation}
G_{\alpha \beta} = 8 \pi T_{\alpha \beta}, \label{Eeq}
\end{equation}
read:

\begin{eqnarray}
8\pi T_{00}=8\pi \tilde \mu
=\left(2\frac{\dot{B}}{B}+\frac{\dot{R}}{R}\right)\frac{\dot{R}}{R}\nonumber\\
-\left(\frac{1}{B}\right)^2\left[2\frac{R^{\prime\prime}}{R}+\left(\frac{R^{\prime}}{R}\right)^2
-2\frac{B^{\prime}}{B}\frac{R^{\prime}}{R}-\left(\frac{B}{R}\right)^2\right],
\label{12}
\end{eqnarray}
\begin{equation}
8\pi T_{01}=-8\pi \tilde q B =-2\left(\frac{{\dot
R}^{\prime}}{R} -\frac{\dot B}{B}\frac{R^{\prime}}{R}\right),
\label{13}
\end{equation}
\begin{eqnarray}
8\pi T_{11}=8\pi \epsilon B^2
=-B^2\left[2\frac{\ddot{R}}{R}+\left(\frac{\dot{R}}{R}\right)^2\right]\nonumber \\
+\left(\frac{R^{\prime}}{R}\right)^2-\left(\frac{B}{R}\right)^2,
\label{14}
\end{eqnarray}
\begin{eqnarray}
8\pi T_{22}=0
=-R^2\left(\frac{\ddot{B}}{B}+\frac{\ddot{R}}{R}
+\frac{\dot{B}}{B}\frac{\dot{R}}{R}\right)\nonumber \\
+\left(\frac{R}{B}\right)^2\left(
\frac{R^{\prime\prime}}{R}
-\frac{B^{\prime}}{B}\frac{R^{\prime}}{R}\right),\label{15}
\end{eqnarray}
where dots and primes denote derivatives with respect to $t$ and $r$ respectively. Observe that  if $\epsilon\neq 0$, the dissipative dust  behaves as an anisotropic  fluid with vanishing tangential stresses \cite{Herreraanis}.

\subsection{Kinematical variables and the mass function}

The expansion $\Theta$ is given by

\begin{equation}
\Theta={V^{\alpha}}_{;\alpha}=\left(\frac{\dot{B}}{B}+2\frac{\dot{R}}{R}\right),\label{th}
\end{equation}
and for the shear we have (remember that the four--acceleration vanishes)

\begin{equation}
\sigma_{\alpha\beta}=V_{(\alpha
;\beta)}-\frac{1}{3}\Theta h_{\alpha\beta}, \label{4a}
\end{equation}
where $h_{\alpha \beta} = g_{\alpha \beta} + V_\alpha V_\beta $.

Using (\ref{5}) we obtain the non--vanishing components
of  (\ref{4a})
\begin{equation}
\sigma_{11}=\frac{2}{3}B^2\sigma, \;\;
\sigma_{22}=\frac{\sigma_{33}}{\sin^2\theta}=-\frac{1}{3}R^2\sigma,
 \label{5a}
\end{equation}
with
\begin{equation}
\sigma^{\alpha\beta}\sigma_{\alpha\beta}=\frac{2}{3}\sigma^2,
\label{5b}
\end{equation}
being
\begin{equation}
\sigma=\left(\frac{\dot{B}}{B}-\frac{\dot{R}}{R}\right).\label{5b1}
\end{equation}
$\sigma_{\alpha\beta}$ may be  also  written as
\begin{equation}
\sigma_{\alpha \beta}= \sigma \left(\chi_\alpha \chi_\beta -
\frac{1}{3} h_{\alpha \beta}\right). \label{sh}
\end{equation}

Next, the mass function $m(t,r)$ introduced by Misner and Sharp \cite{Misner} (see also \cite{Cahill})
is given by

\begin{equation}
m=\frac{(R)^3}{2}{R_{23}}^{23}=\frac{R}{2}\left[\dot R^2-\left(\frac{R^{\prime}}{B}\right)^2+1   \right].
\label{18}
\end{equation}

We can define the velocity $U$ of the collapsing
fluid as the variation of the areal  radius ($R$) with respect to proper time, i.e.\
\begin{equation}
U=\dot R. \label{19}
\end{equation}
Then (\ref{18}) can be rewritten as
\begin{equation}
E \equiv \frac{R^{\prime}}{B}=\left[1+U^2-\frac{2m(t,r)}{R}\right]^{1/2}.
\label{20}
\end{equation}
With the above  we can express (\ref{13}) as
\begin{equation}
4\pi \tilde q=E\left[\frac{1}{3 R^{\prime}}(\Theta-\sigma)^{\prime}
-\frac{\sigma}{R}\right].\label{21a}
\end{equation}

From
(\ref{18})
\begin{eqnarray}
\dot m=-4\pi(
\epsilon U+\tilde q E)R^2,
\label{22}
\end{eqnarray}
and
\begin{eqnarray}
m^{\prime}=4\pi\left(\tilde \mu+\tilde q \frac{U}{E}\right)R^{\prime}R^2.
\label{27}
\end{eqnarray}
Equation (\ref{27}) may be integrated to obtain
\begin{equation}
m=\int^{r}_{0}4\pi R^2 \left(\tilde \mu
+\tilde q\frac{U}{E}\right)R^{\prime}dr\label{27int}
\end{equation}
(assuming a regular centre to the distribution, so $m(0)=0$). We may partially integrate (\ref{27int}) to obtain
\begin{equation}
\frac{3m}{R^3} = 4\pi\tilde{\mu} - \frac{4\pi}{R^3} \int^r_0{R^3\left({\tilde \mu^{\prime}}-3 \tilde q \frac{R^{\prime}U}{RE}\right)dr}.
\label{3m/R3}
\end{equation}

\subsection{The exterior spacetime and junction conditions}
In the case of bounded configurations,  we assume that outside $\Sigma$ we have the Vaidya
spacetime (i.e.\ we assume all outgoing radiation is massless), or in the dissipationless case the Schwarzschild spacetime,
described by
\begin{equation}
ds^2=-\left[1-\frac{2M(v)}{\rho}\right]dv^2-2d\rho dv+\rho^2(d\theta^2
+\sin^2\theta
d\phi^2) \label{1int},
\end{equation}
where $M(v)$  denotes the total mass (which is constant in the Schwarzschild case) and  $v$ is the retarded time.

The matching of the full nonadiabatic sphere  (including viscosity) to
the Vaidya spacetime, on the surface $r=r_{\Sigma}=$ constant, was discussed in
\cite{chan1} (for the discussion of the shear--free case see \cite{Santos} and \cite{Bonnor}). However observe that we are now including  a null fluid within the star configuration.

Now, from the continuity of the first  differential form it follows (see \cite{chan1} for details),
\begin{equation}
 dt\stackrel{\Sigma}{=}dv \left(1-\frac{2M(v)}{\rho}\right), \label{junction1f}
\end{equation}
\begin{equation}
R\stackrel{\Sigma}{=}\rho(v), \label{junction1f2}
\end{equation}
and
 \begin{equation}
\left(\frac{dv}{dt}\right)^{-2}\stackrel{\Sigma}{=}\left(1-\frac{2M(v)}{\rho}+2\frac{d\rho}{dv}\right).\label{junction1f3}
\end{equation}

Whereas the continuity of the second differential form produces

\begin{equation}
m(t,r)\stackrel{\Sigma}{=}M(v), \label{junction1}
\end{equation}
and
\begin{widetext}
\begin{eqnarray}
2\left(\frac{{\dot R}^{\prime}}{R}-\frac{\dot B}{B}\frac{R^{\prime}}{R}\right)
\stackrel{\Sigma}{=}-B\left[2\frac{\ddot R}{R}+\left(\frac{\dot R}{R}\right)^2\right]+\frac{1}{B}\left[\left(\frac{R^{\prime}}{R}\right)^2-\left(\frac{B}{R}\right)^2\right],
\label{j2}
\end{eqnarray}
\end{widetext}
where $\stackrel{\Sigma}{=}$ means that both sides of the equation
are evaluated on $\Sigma$ (observe a misprint in eq.(40) in \cite{chan1} and a slight difference in notation).

Comparing (\ref{j2}) with  (\ref{13}) and (\ref{14}) one obtains
\begin{equation}
q\stackrel{\Sigma}{=}0.\label{j3}
\end{equation}
Thus   the matching of
(\ref{1})  and (\ref{1int}) on $\Sigma$ implies (\ref{junction1}) and  (\ref{j3}).

Also, we have
\begin{equation}
\epsilon \stackrel{\Sigma}{=}\frac{L}{4\pi \rho^2}, \label{20lum}
\end{equation}
where (\ref{j3}) has been used and  $L_\Sigma$ denotes   the total luminosity of the  sphere as measured on its surface, which is given by
\begin{equation}
L \stackrel{\Sigma}{=}L_{\infty}\left(1-\frac{2m}{\rho}+2\frac{d\rho}{dv}\right)^{-1}, \label{14a}
\end{equation}
and where
\begin{equation}
L_{\infty} =\frac{dM}{dv}\stackrel{\Sigma}{=} -\left[\frac{dm}{dt}\left(\frac{dv}{dt}\right)^{-1}\right],\label{14b}
\end{equation}
is the total luminosity measured by an observer at rest at infinity.

The boundary redshift $z_\Sigma$ is given by
\begin{equation}
\frac{dv}{dt}\stackrel{\Sigma}{=}1+z,
\label{15b}
\end{equation}
with
\begin{equation}
\frac{dv}{dt}\stackrel{\Sigma}{=}\left(\frac{R^{\prime}}{B}+\dot R\right)^{-1}.
\label{16b}
\end{equation}
Therefore the time of formation of the black hole is given by
\begin{equation}
\left(\frac{R^{\prime}}{B}+\dot R\right)\stackrel{\Sigma}{=}E+U\stackrel{\Sigma}{=}0.
\label{17b}
\end{equation}
Also observe than from (\ref{junction1f3}), (\ref{14a})  and (\ref{16b}) it follows
\begin{equation}
L\stackrel{\Sigma}{=}\frac{L_\infty}{(E+U)^2},
\label{ju}
\end{equation}
and from (\ref{19}), (\ref{20}), (\ref{junction1f3}) and (\ref{16b})
\begin{equation}
\frac{d\rho}{dv}\stackrel{\Sigma}{=}U(U+E).
\label{juf}
\end{equation}

Finally it is worth noticing that in the diffussion approximation ($\epsilon=0, q\neq0$) it follows at once from (\ref{20lum}) that the total luminosity ($L_\Sigma$) vanishes, even though there are  non--vanishing dissipative fluxes ($q$) within the sphere. This result is an obvious consequence of the dust condition (vanishing hydrodynamic pressure).

\subsection{ Weyl tensor}

The Weyl tensor is defined through the  Riemann tensor
$R^{\rho}_{\alpha \beta \mu}$, the  Ricci tensor
$R_{\alpha\beta}$ and the curvature scalar $\cal R$, as:
$$
C^{\rho}_{\alpha \beta \mu}=R^\rho_{\alpha \beta \mu}-\frac{1}{2}
R^\rho_{\beta}g_{\alpha \mu}+\frac{1}{2}R_{\alpha \beta}\delta
^\rho_{\mu}-\frac{1}{2}R_{\alpha \mu}\delta^\rho_\beta$$
\begin{equation}
+\frac{1}{2}R^\rho_\mu g_{\alpha \beta}+\frac{1}{6}{\cal
R}(\delta^\rho_\beta g_{\alpha \mu}-g_{\alpha
\beta}\delta^\rho_\mu). \label{34}
\end{equation}

The electric  part of  Weyl tensor is defined by
\begin{equation}
E_{\alpha \beta} = C_{\alpha \mu \beta \nu} V^\mu V^\nu,
\label{elec}
\end{equation}
with the following non--vanishing components
\begin{eqnarray}
E_{11}&=&\frac{2}{3}B^2 {\cal E},\nonumber \\
E_{22}&=&-\frac{1}{3} R^2 {\cal E}, \nonumber \\
E_{33}&=& E_{22} \sin^2{\theta}, \label{ecomp}
\end{eqnarray}
where
\begin{eqnarray}
&&{\cal E}= \frac{1}{2}\left[\frac{\ddot R}{R} - \frac{\ddot B}{B} - \left(\frac{\dot R}{R} - \frac{\dot B}{B}\right) \frac{\dot R}{R}\right]\nonumber \\
&+& \frac{1}{2 B^2} \left[ -
\frac{R^{\prime\prime}}{R} + \left(\frac{B^{\prime}}{B} +
\frac{R^{\prime}}{R}\right) \frac{R^{\prime}}{R}\right]
-\frac{1}{2 R^2}. \label{E}
\end{eqnarray}

Observe that we may also write $E_{\alpha\beta}$ as:
\begin{equation}
E_{\alpha \beta}={\cal E} (\chi_\alpha
\chi_\beta-\frac{1}{3}h_{\alpha \beta}). \label{52}
\end{equation}

Finally, using (\ref{12}), (\ref{14}), (\ref{15}) with (\ref{18}) and (\ref{E}) we obtain
\begin{equation}
\frac{3m}{R^3}=4\pi \left(\tilde{\mu}-\epsilon \right) - \cal{E}.
\label{mE}
\end{equation}

\subsection{Evolution equations for the expansion and the shear,   Bianchi identities and a constraint equation  for the Weyl tensor}
For the system under consideration two equations describing the evolution of the expansion (Raychaudhury) and the shear (Ellis \cite{ellis1}, \cite{ellis2}), respectively,  can be easily derived (see \cite{she} for details), they read:
\begin{equation}
\dot{\Theta}+\frac{1}{3}\Theta ^2+\frac{2}{3}\sigma
^2=-4\pi(\tilde
\mu+\epsilon), \label{c}
\end{equation}

\begin{equation}
\dot \sigma+\frac{1}{3}\sigma
^2+\frac{2}{3}\Theta\sigma=4\pi\epsilon-{\cal E}.\label{d}
\end{equation}

The two independent components of Bianchi identities for the system under consideration read (see \cite{H1} for details):
\begin{eqnarray}
\dot{\tilde \mu}+(\tilde \mu +\epsilon)\frac{\dot B}{B}+2\frac{\tilde
\mu \dot R}{R}+\frac{\tilde
q^\prime}{B}+2\frac{\tilde q R^\prime}{BR}=0,\label{an}
\end{eqnarray}
\\
\begin{eqnarray}
\dot{\tilde q}+
\frac{\epsilon^\prime}{B}+2\tilde q\left(\frac{\dot B}{B}+\frac{\dot R}{R}\right)+2\frac{R^\prime \epsilon}{R B}=0.\label{bn}
\end{eqnarray}

Finally, the following constraint equation for the Weyl tensor may be derived from the Bianchi identities (e.g. see \cite{Hetal} for details)
\begin{equation}
\left[{{\cal E}}-4\pi(\tilde
\mu-\epsilon)\right]^\prime+({\cal E}+4\pi \epsilon)\frac{3
R^\prime}{R}=-12\pi\tilde q B\frac{\dot R}{R}. \label{h}
\end{equation}

\section{Structure Scalars}
As we mentioned in the Introduction, part of our analysis will be based on a set of scalar functions which   appear in a natural way in the orthogonal splitting of the Riemann tensor (see \cite{sp} for details).

Thus, let us introduce  the tensors $Y_{\alpha \beta}$ and
$X_{\alpha \beta}$ (which are  elements of that splitting \cite{bel1}, \cite{parrado}), defined
by:
\begin{equation}
Y_{\alpha \beta}=R_{\alpha \gamma \beta \delta}V^\gamma V^\delta,
\label{electric}
\end{equation}
and
\begin{equation}
X_{\alpha \beta}=^*R^{*}_{\alpha \gamma \beta \delta}V^\gamma
V^\delta=\frac{1}{2}\eta_{\alpha\gamma}^{\quad \epsilon
\rho}R^{*}_{\epsilon \rho\beta\delta}V^\gamma V^\delta,
\label{magnetic}
\end{equation}

\noindent where $R^*_{\alpha \beta \gamma \delta}=\frac{1}{2}\eta
_{\epsilon \rho \gamma \delta} R_{\alpha \beta}^{\quad \epsilon
\rho}$ and $\eta
_{\epsilon \rho \gamma \delta}$ denotes the Levi--Civita tensor.

\noindent Tensors $Y_{\alpha \beta}$ and  $X_{\alpha \beta}$ may also be expressed through their traces and their trace--free parts, as
\begin{eqnarray}
Y_{\alpha\beta}=\frac{1}{3}Y_T h_{\alpha
\beta}+Y_{TF}(\chi_{\alpha} \chi_{\beta}-\frac{1}{3}h_{\alpha
\beta}), \label{electric'}
\\
X_{\alpha\beta}=\frac{1}{3}X_T h_{\alpha
\beta}+X_{TF}(\chi_{\alpha} \chi_{\beta}-\frac{1}{3}h_{\alpha
\beta}).\label{magnetic'}
\end{eqnarray}

Then from  (\ref{electric})-(\ref{magnetic'}) and using
(\ref{Tab}), (\ref{Eeq}), (\ref{34}) and  (\ref{52}), we obtain

\begin{equation}
Y_T=4\pi(\tilde \mu+\epsilon), \qquad
Y_{TF}={\cal E}-4\pi\epsilon, \label{EY}
\end{equation}

\begin{equation}
X_T=8\pi\tilde \mu , \qquad
X_{TF}=-{\cal E}-4\pi \epsilon. \label{EX}
\end{equation}
Also, combining (\ref{3m/R3}) and (\ref{mE}) with (\ref{EY}) we may write
\begin{equation}
Y_{TF}= -8\pi\epsilon +\frac{4\pi}{R^3}\int^r_0{R^3\left( \tilde{\mu}^{\prime}-3\tilde{q}\frac{U R^{\prime}}{RE}\right)dr}.
\label{Y}
\end{equation}

Thus the scalar $Y_{TF}$ may be expressed through the Weyl tensor and the outgoing null fluid   or in terms of the null radiation, the density inhomogeneity and  the dissipative variables. It is worth recalling that a link between $Y_{TF}$ and the Tolman mass has been established in \cite{sp}. Also, it has been shown that in the geodesic case, $Y_{TF}$ controls the stability of the shearfree condition \cite{she}.

Scalars (\ref{EY})-(\ref{EX}) for the line element  (\ref{1}) are
\begin{equation}
Y_T=-2\frac{\ddot R}{R}-\frac{\ddot
B}{B}, \qquad
Y_{TF}=\frac{\ddot R}{R}-\frac{\ddot
B}{B},\label{YTFM}
\end{equation}

\begin{eqnarray}
X_T&=&\left(2\frac{\dot{B}}{B}+\frac{\dot{R}}{R}\right)\frac{\dot{R}}{R}\nonumber \\
&-&\frac{1}{B^2}\left[2\frac{R^{\prime\prime}}{R}+\left(\frac{R^{\prime}}{R}\right)^2
-2\frac{B^{\prime}}{B}\frac{R^{\prime}}{R}\right]+\frac{1}{R^2},
\label{XT}
\end{eqnarray}

\begin{eqnarray}
X_{TF}&=&\left( \frac{\dot R}{R}-\frac{\dot
B}{B}\right)\frac{\dot R}{R}\nonumber \\&+&\frac{1}{B^2}\left[\frac{R^{\prime
\prime}}{R}-\left(\frac{B^\prime}{B}+\frac{R^\prime}{R}\right)\frac{R^\prime}{R}\right]+\frac{1}{R^2}.\label{XTFM}
\end{eqnarray}

In terms of these scalar functions, equations (\ref{c}), (\ref{d}) and (\ref{h}), become
\begin{equation}
\dot{\Theta}+\frac{1}{3}\Theta ^2+\frac{2}{3}\sigma
^2=-Y_T, \label{c'}
\end{equation}
\\
\begin{equation}
\dot \sigma+\frac{1}{3}\sigma
^2+\frac{2}{3}\Theta\sigma=-Y_{TF},\label{d'}
\end{equation}
\begin{equation}
(4\pi\tilde
\mu+X_{TF})^\prime=-\frac{3R^\prime}{R}X_{TF}+4\pi\tilde
q(\Theta-\sigma)B.
\label{h'}
\end{equation}
Now, from (\ref{h'}) in the non--dissipative case we have
\begin{equation}
(4\pi
\mu+X_{TF})^\prime=-\frac{3R^\prime}{R}X_{TF},
\label{h''}
\end{equation}
implying that if  $X_{TF}=0$ then $\mu^{\prime}=0$. On the other hand if $\mu^{\prime}=0$ then
\begin{equation}
(X_{TF})^\prime=-\frac{3R^\prime}{R}X_{TF},
\label{h1''}
\end{equation}
producing
\begin{equation}
X_{TF}=\frac{f(t)}{R^3}.
\label{h2''}
\end{equation}
Since $X_{TF}$ must be regular everywhere within the sphere, we must put $f(t)=0$, therefore we have
$\mu^{\prime}=0$ $\Longleftrightarrow$ $X_{TF}=0$. In other words, in absence of dissipation, energy density inhomogeneity is controled by the scalar $X_{TF}$. This last result is also valid for an anisotropic fluid  \cite{sp}.

In the case of non--dissipative  dust, it can be shown that $\sigma=0$ implies $X_{TF}=0$.
Indeed, in this particular case it follows from (\ref{d'}) that $Y_{TF}=0$, which because of (\ref{EY}) and (\ref{EX})  produces $X_{TF}=0$.

We shall next analyze in some detail some properties  of LTB spacetimes.

\section{Lemaitre--Tolman--Bondi metric}
In order to obtain the general form of LTB, we assume a geodesic, dissipationless fluid ($q=\epsilon=0$). Then
we find after integration of  (\ref{13}),
\begin{equation}
B(t,r)=\frac{R^\prime}{(1+\kappa(r))^{1/2}},\label{BTB}
\end{equation}
where $\kappa$ is an arbitrary function of $r$.

Feeding back  (\ref{BTB}) in (\ref{1}) we obtain the Lemaitre-Tolman-Bondi metric (LTB):
\begin{equation}
ds^2=-dt^2+\frac{(R^{\prime})^2}{1+\kappa(r)}dr^2+R^2(d\theta^2+\sin^2\theta
d\phi^2).\label{mTB}
\end{equation}
The metric (\ref{mTB}) is usually associated with an inhomogeneous  dust source (and so we shall do here), however it is worth mentioning that  the most general source compatible with LTB spacetimes is an anisotropic fluid (\cite{4}, \cite{s1}).

Next, the ``Euclidean'' condition at the centre requires that the perimeter of an infinitesimally small circle around the centre be given by $2 \pi l$
where $l$ is the proper radius of the circle, given in turn by
 \begin{equation}
l=\int_0^r Bdr.
\label{ml}
\end{equation}
On the other hand, the perimeter of any circle, as it follows from (\ref{1}) is given by $ 2 \pi R$
implying that in the neighborhood of the centre we must impose the condition:
\begin{equation}
R^{\prime\stackrel{r=0}{=}}B,
\label{nc0}
\end{equation}
or, using (\ref{BTB})
\begin{equation}
\kappa\stackrel{r=0}{=}0.
\label{nc01}
\end{equation}

Now from (\ref{18}) and (\ref{BTB}) we obtain
\begin{equation}
\dot R^2=\frac{2m}{R}+\kappa(r),
\label{intltb1}
\end{equation}
and from (\ref{22}) and  (\ref{27})
\begin{equation}
m=m(r),
\label{int3}
\end{equation}
\begin{equation}
m^{\prime}=4\pi\mu R^2R^{\prime}.
\label{int2}
\end{equation}

Also,  the Bianchi identity (\ref{an}) in this case reads

\begin{eqnarray}
\dot \mu+\mu\left(\frac{\dot B}{B}+2\frac{\dot R}{R}\right)=0,\label{an'}
\end{eqnarray}
producing
\begin{equation}
 \mu=\frac{h(r)}{BR^2},\label{nltb}
\end{equation}
or, using (\ref{BTB})
\begin{equation}
 \mu=\frac{3 h(r)(1+\kappa(r))^{1/2}}{(R^{3})^\prime},\label{n1ltb}
\end{equation}
where $h(r)$ is a function of integration.

Depending on $\kappa$ there are three possible solutions of (\ref{intltb1}):
\begin{enumerate}
\item $\kappa=0$ (Parabolic case)
\begin{equation}
 R^{3/2}=(2m)^{1/2}\eta^3, \qquad \frac{2}{3}\eta^3=t-t_{bb}(r).
\label{int4}
\end{equation}
\item $\kappa>0$ (Hyperbolic case)
\begin{equation}
 R=\frac{m}{\kappa}(\cosh \eta-1), \qquad  \frac{m}{\kappa^{3/2}}(\sinh \eta-\eta)=t-t_{bb}(r).
\label{int5}
\end{equation}
\item $\kappa<0$ (Elliptic case)
\begin{equation}
 R=\frac{m}{|\kappa|}(1-\cos \eta), \qquad  \frac{m}{|\kappa|^{3/2}}(\eta-\sin \eta)=t-t_{bb}(r).
\label{int5e}
\end{equation}
Where $t_{bb}(r)$ is an integration function of $r$ giving the value of time for which $R(t,r)=0$, not to confound with the center of symmetry $R(t,0)$.
\end{enumerate}
In order to prescribe an explicit model we have to provide the three functions $\kappa(r)$, $m(r)$ and $t_{bb}(r)$. However, since (\ref{mTB}) is invariant under  transformations of the form $r=r(\tilde r)$, we only need two functions of $r$.

\noindent Scalars $Y_T$,
 $Y_{TF}$, $X_{TF}$ and $X_T$ for (\ref{mTB}) are:
\begin{equation}
Y_{TF}=\frac{\ddot R}{R}-\frac{\ddot R^\prime}{R^\prime},\qquad
Y_T=-2\frac{\ddot R}{R}-\frac{\ddot
R^\prime}{R^\prime},\label{EYTB}
\end{equation}
and
\begin{equation}
X_{TF}=\frac{\dot R^2}{R^2}-\frac{\dot R}{R}\frac{\dot R^\prime}{R^\prime}-\frac{\kappa}{R^2}+\frac{\kappa^{\prime}}{2RR^{\prime}},\label{EXTFB}
\end{equation}
\begin{equation}
X_T=\frac{2\dot R \dot R^\prime}{R R^{\prime}}+\frac{\dot R^2}{R^2}-\frac{\kappa}{R^2}-\frac{\kappa^{\prime}}{RR^{\prime}}.\label{EXTB}
\end{equation}

\subsection{Symmetry properties of LTB spacetimes}
It is a simple matter to check that besides the Killing vectors associated with spherical symmetry, LTB spacetimes admit no further Killing vectors. Also, it is not difficult to verify that they do not admit conformal Killing vectors either.  It should be clear that  such nonadmittance refers to the general form of LTB, not to any specific  solution belonging to LTB spacetime.

However  as we shall see below, a specific kind of symmetry may be ascribed to LTB spacetimes.

\subsubsection{Proper matter collineation}

It can be shown \cite{Carot94} that in the pure dust case, there exists a vector field $\xi$ such that  for any LTB spacetime

\begin{equation}
\pounds_ \xi T_{\alpha \beta}=0,
\label{li0}
\end{equation}
where  $\pounds_ \xi $ denotes the Lie derivative with respect to the vector field ${\bf \xi}$, which can be shown to be of the form

\begin{equation}
\xi^\rho = \xi^0 \delta_t^\rho + \xi^1 \delta_r^\rho.
\label{kid}
\end{equation}
whose components are are given by
\begin{eqnarray}
\xi^0 &=& F(t) \label{xi0},\\
\xi^1 &=& -\left(2 \dot F(t) + F(t) \frac{\dot \mu}{\mu}\right)\left(\frac{\mu}{\mu^\prime}\right),
\label{xi1}
\end{eqnarray}
where $F(t)$ is an arbitrary function. Thus for any LTB spacetime we can always  find a vector field $\xi$ satisfying (\ref{li0}).

It should be stressed that $\xi$ is {\it not} a Killing vector field, i.e. it defines a proper matter collineation.

\section{Generalizing LTB to the dissipative case}

From the previous section it should be clear that LTB spacetimes explicitly exclude dissipative fluxes (either  in the diffusion approximation or in the streaming out limit).

We shall now  approach the problem of extending LTB so as to include dissipative fluxes (such spacetimes will be  referred to as GLTB). As a necessary condition we shall  require that all GLTB's become LTB in the limit when dissipative fluxes vanish.

Since we are searching for spacetimes  as ``close'' as possible to LTB,  we shall  consider geodesic, shearing,  dust with  $ \tilde q\neq 0$. It is worth recalling that   the pure dust (without dissipation) condition implies that the fluid is geodesic, however this is no longer true for the  dissipative  case. Therefore the geodesic condition here is non--redundant.

Then integration of (\ref{13}), produces now:
\begin{equation}
B(t,r)=\frac{R^\prime}{\left(1+K(t,r)\right)^{1/2}},\label{HDO}
\end{equation}
\noindent with
\begin{equation}
1+K(t,r)=\left[\int 4\pi\tilde q Rdt+ C(r)\right]^2.\label{Kq}
\end{equation}
Since  in the non-dissipative case  (\ref{HDO}) should become (\ref{BTB}), it follows
\begin{equation}
C(r)=\left(1+\kappa(r)\right)^{1/2},\label{Kq1}
\end{equation}
and the Euclidean condition at the center, implies
\begin{equation}
C(r)\stackrel{r=0}{=}1.
\label{nc02}
\end{equation}

Feeding back (\ref{HDO}) into (\ref{1}) and using (\ref{Kq}),  we obtain for the line element

\begin{eqnarray}
ds^2&=&-dt^2+\frac{(R^{\prime})^2}{\left[\int 4\pi\tilde q Rdt+
C(r)\right]^2}dr^2\nonumber \\&+&R^2(d\theta^2+\sin^2\theta d\phi^2),\label{mHDO}
\end{eqnarray}
or, using  (\ref{Kq1})
\begin{eqnarray}
ds^2&=&-dt^2+\frac{(R^{\prime})^2}{\left[\int 4\pi\tilde q Rdt+
(1+\kappa(r))^{1/2}\right]^2}dr^2\nonumber \\&+&R^2(d\theta^2+\sin^2\theta d\phi^2).\label{mHDO1}
\end{eqnarray}

\noindent Scalars $Y_T$,  $Y_{TF}$, $X_T$ and $X_{TF}$ for
(\ref{mHDO}) are:
\begin{eqnarray}
Y_{TF}&=&\frac{\ddot R}{R}-\frac{\ddot
R^\prime}{R^\prime}+\frac{\dot K}{1+K}\left(\frac{\dot
R^\prime}{R^\prime}-\frac{3}{4}\frac{\dot K}{1+K}\right)\nonumber \\&+&\frac{\ddot
K}{2(1+K)},\label{1sd}
\end{eqnarray}
\begin{eqnarray}
Y_T&=&-2\frac{\ddot R}{R}-\frac{\ddot R^\prime}{R^\prime}+\frac{\dot
K}{1+K}\left(\frac{\dot R^\prime}{R^\prime}-\frac{3}{4}\frac{\dot
K}{1+K}\right)\nonumber \\&+&\frac{\ddot K}{2(1+K)},\label{2sd}
\end{eqnarray}
\begin{eqnarray}
X_T=\frac{2\dot R \dot R^\prime}{R R^{\prime}}+\frac{\dot R^2}{R^2}-\frac{K}{R^2}-\frac{K^{\prime}}{RR^{\prime}}-\frac{\dot R \dot K}{R(1+K)},\label{3sd}
\end{eqnarray}
\begin{eqnarray}
X_{TF}=\frac{\dot R^2}{R^2}-\frac{\dot R}{R}\frac{\dot R^\prime}{R^\prime}-\frac{K}{R^2}+\frac{K^{\prime}}{2RR^{\prime}}+\frac{\dot R\dot K}{2R(1+K)}.
\label{EYHDO}
\end{eqnarray}

In order to proceed further and to obtain specific families of solutions we need to impose additional conditions. As mentioned before the criteria to select such conditions will be dictated by the requirement that the obtained solution represents the ``closest'' possible situation to LTB spacetime, including dissipative fluxes.

Of course, this last requirement is still  vague enough and  allows a great deal of possibilities. In this work we shall focus on two possible extensions. On the one hand we shall consider  extensions based on the characterization of LTB spacetimes  in terms of the  structure scalars. On the other hand  we shall propose extensions of LTB to the dissipative case, based on the symmetry property discussed above. However before considering that, let us describe the treatment of the transport equation in the pure diffusive case.

\section{THE TRANSPORT EQUATION}
 In the diffusion approximation ($\epsilon=0, \tilde q=q$), we shall use a transport equation derived from the
M\"{u}ller-Israel-Stewart second
order phenomenological theory for dissipative fluids \cite{Muller67, IsSt76, I, II}.

Indeed, it is well known that the Maxwell-Fourier law for
radiation flux leads to a parabolic equation (diffusion equation)
which predicts propagation of perturbations with infinite speed
(see \cite{6D}-\cite{8'} and references therein). This simple fact
is at the origin of the pathologies \cite{9H} found in the
approaches of Eckart \cite{10E} and Landau \cite{11L} for
relativistic dissipative processes. To overcome such difficulties,
various relativistic
theories with non-vanishing relaxation times have been proposed in
the past \cite{Muller67,IsSt76, I, II, 14Di,15d}. The important point is that
all these theories provide a heat transport equation which is not
of Maxwell-Fourier type but of Cattaneo type \cite{18D}, leading
thereby to a hyperbolic equation for the propagation of thermal
perturbations.

The corresponding  transport equation for the heat flux reads
\begin{equation}
\beta
h^{\alpha\rho}V^{\gamma}q_{\rho;\gamma}+q^{\alpha}=-K h^{\alpha\rho}
({\cal T}_{,\rho}+{\cal T}a_{\rho}) -\frac 12\kappa {\cal T}^2\left( \frac{\beta
V^\rho }{K }{{\cal T}^2}\right) _{;\rho }q^\alpha ,  \label{21t}
\end{equation}
where $K $  denotes the thermal conductivity, and  ${\cal T}$ and
$\beta$ denote temperature and relaxation time respectively. 
In the case $\beta=0$ we recover the Eckart--Landau equation.

Observe
that, due to the symmetry of the problem, equation (\ref{21t}) only
has one independent component, which may be written  as
\begin{equation}
\beta \dot{q}+q =-\frac{K}{B} {\cal T}^{\prime}
-\frac{1}{2}\beta  q \Theta+\beta\frac{\dot {\cal T}}{{\cal T}} q.
\label{ISE'}
\end{equation}
In the truncated version of the theory, the last term in (\ref{21t}) is absent (see for example \cite{pavon}), and (\ref{ISE'}) becomes
\begin{equation}
\beta \dot{q}+q =-\frac{K}{B} {\cal T}^{\prime}.
\label{ISE'bis}
\end{equation}

Finally observe that if $\epsilon = 0$ we obtain from  (\ref{bn})
\begin{equation}
q=\frac{g(r)}{B^2R^2},
\label{difn1}
\end{equation}
where $g(r)$ is an arbitrary function.
Then using (\ref{difn1}) in (\ref{ISE'}) or (\ref{ISE'bis}) we obtain (up to the function $g(r)$) the temperature distribution of the fluid in terms of metric functions.

\section{EXTENSIONS OF LTB BASED ON STRUCTURE SCALARS}

\noindent From an inspection of  (\ref{c'}), (\ref{d'}), (\ref{EYTB})--(\ref{EXTB}) and (\ref{1sd})--(\ref{EYHDO}), the following remarks are in order:
\begin{itemize}
\item   For any geodesic fluid (dissipative or not),
the evolution of $\Theta$ and $\sigma$ is fully controlled by $Y_{TF}$ and $Y_T$
\item In LTB, scalars $Y_T$ and $Y_{TF}$ do not contain $\kappa$.
\item  In GLTB, scalars $Y_T$ and $Y_{TF}$ differ from their expressions in LTB, by the same term.
\end{itemize}

Based on the comments above and on the requirement  of ``maximal'' similarity between a LTB and its  corresponding GLTB,  let us assume that scalars  $Y_{T}$ and  $Y_{TF}$ share  the same expression (in terms of $R$) in both cases.

\noindent Then it follows at once from (\ref{EYTB})  and
(\ref{1sd}) (or (\ref{2sd})),
\begin{equation}
\frac{\dot K}{1+K}\left[\frac{\dot
R^\prime}{R^\prime}-\frac{3}{4}\frac{\dot
K}{(1+K)}\right]+\frac{\ddot K}{2(1+K)}=0.\label{EK}
\end{equation}

\noindent Integrating  (\ref{EK}) we obtain
\begin{equation}
\frac{R^\prime \dot K^{1/2}}{(1+K)^{3/4}} = C_1(r),
\label{1aInt}
\end{equation}
or using (\ref{Kq}) and (\ref{Kq1})
\begin{equation}
C_1(r)=\frac{(8\pi \tilde q R)^{1/2} R^\prime  }{(1+\kappa(r))^{1/2}+\int{4 \pi \tilde q R dt}},
\label{c1}
\end{equation}
where $C_1(r)$ is an arbitrary integration function.

Next, integrating (\ref{1aInt}) we obtain
\begin{equation}
K+1=\frac{4}{ \left[C_1(r)^2\int{\frac{dt}{(R^\prime)^2}+C_2 (r)}\right]^{2}},
\label{2aInt}
\end{equation}
where $C_2(r)$ is another integration function, which may be related to $C(r)$
as follows.
Taking the $t$ derivative of  (\ref{2aInt}) we have
\begin{equation}
\dot K=-\frac{8C_1^2(r)}{(R^\prime)^2\left[C_1^2(r)\int
\frac{dt}{(R^\prime) ^2}+C_2(r)\right]^3},\label{KP}
\end{equation}

\noindent then, combining  (\ref{Kq}) with (\ref{KP}) it follows
\begin{equation}
2\pi\tilde q=\frac{C_1^2(r)}{R (R^{\prime})^2\left[C_1^2(r)\int
\frac{dt}{(R^\prime) ^2}+C_2(r)\right]^2},
\label{qR}
\end{equation}
finally, comparing (\ref{c1}) and (\ref{qR}) we also find
\begin{equation}
C_2(r) = \frac{2\left(1-4\pi \tilde q R R^{\prime2}\right)}{C(r) + \int{4 \pi \tilde q R dt}}.
\label{c2}
\end{equation}

\noindent Thus, the assumption above on $Y_T$ and $Y_{TF}$ allows us to obtain a GLTB from any given LTB. Indeed, from (\ref{qR}), taking the time and radial dependence of $R$ from any given LTB, we obtain the dissipative flux of the corresponding GLTB, up to the two functions of $r$, $C_1$ and $C_2$. The former must satisfy the regularity condition $C_1(0)=0$, and its vanishing brings back to the starting ``seed'' LTB solution. The remaining physical variables follow from the field equations.

Let us now illustrate the method above, with an example inspired in the parabolic subclass of LTB

Thus, we choose:
\begin{equation}
R(t,r)=f(r)(T(r)-t)^{2/3}.\label{RT}
\end{equation}
where $f(r)$ and $T(r)$ correspond to any specific parabolic LTB solution. However, care must be exercised in not identifying $f(r)$ with the mass function as it happens in LTB.

\noindent Using  (\ref{RT}) we may write the integral  $\int
\frac{dt}{(R^{\prime})^2}$ as:
\begin{equation}
I=\int
\frac{dt}{(R^{\prime})^2}=\int\frac{(T(r)-t)^{2/3}dt}{\left[f^{\prime}(r)(T(r)-t)+\frac{2}{3}f(r)T^{\prime}
(r)\right] ^2},\label{i}
\end{equation}

\noindent introducing the variable $T(r)-t=\tau^3$ in  (\ref{i}) and integrating we obtain:
$$
I=-3\int\frac{\tau ^4 d\tau}{(a\tau ^3+b)^2}=
$$
\begin{widetext}
\begin{eqnarray}
=\frac{\tau^2}{(a^2 \tau^3+ab)}+\frac{1}{3\sqrt[3]{ba^5}}\left[2
\sqrt{3}\arctan\left(\frac{1-2\sqrt[3]{\frac{a}{b}}\tau}{\sqrt{3}}\right)-\ln
\left(\frac{a\tau^3+b}{(\sqrt[3]{a} \tau +\sqrt[3]{b})^3}\right)\right],
\label{ii}
\end{eqnarray}
\end{widetext}
\noindent where $a=f^{\prime}(r)$ and  $b=\frac{2}{3}f(r)
T^{\prime}(r)$.

Feeding back  (\ref{ii}) into  (\ref{qR}), using
(\ref{RT}), we find:
\begin{widetext}
\begin{equation}
2\pi \tilde q= \frac{C_1^2(r)f(r)^{-1}}{\left\{
C_1^2(r)\left[\frac{\tau^2}{a}+\frac{a\tau^3+b}{3\sqrt[3]{ba^5}}\left[2
\sqrt{3}\arctan\left(\frac{1-2\sqrt[3]{\frac{a}{b}}\tau}{\sqrt{3}}\right)-\ln
\left(\frac{a\tau^3+b}{(\sqrt[3]{a} \tau
+\sqrt[3]{b})^3}\right)\right]\right]+C_2(r)(a\tau^3+b)\right\}^2}.\label{qRT}
\end{equation}
\end{widetext}

Thus, the  function $C_1$ controls the magnitude of the  dissipation and can be chosen as small  as desired in the case when perturbations of LTB are required. From the above it is evident that these GLTB become LTB when dissipative fluxes vanish. Also, it should be clear that if the  physical properties of the ``seed'' LTB  are reasonable, so will be the properties of the corresponding GLTB, at least for a sufficiently small value of the function $C_1(r)$. Finally, in the particular case ($\epsilon=0$) the temperature profile of the model can be obtained from (\ref{ISE'}) (or (\ref{ISE'bis})) and (\ref{qRT}).

\section{EXTENSIONS OF LTB BASED ON SYMMETRY PROPERTIES}

In this section we shall propose another approach to obtain GLTB spacetimes. It consists in assuming that any GLTB  spacetime shares the same symmetry property described in section IV for the corresponding LTB.

Thus, let us  assume now, that  our GLTB admits a proper matter collineation.

Then, from the energy momentum tensor for  dust with dissipation (\ref{Tab}), we obtain, using (\ref{li0}),  the  three equations:
\begin{equation}
\xi^0 \dot{ \tilde \mu} + \xi^1 \tilde \mu^\prime+2\tilde\mu \xi^0_{,0}- 2\tilde q B \xi^1_{,0} =0,
\label{eno1}
\end{equation}
\begin{equation}
-\xi^0 \dot {\tilde q} - \xi^1 {\tilde q}^\prime +\tilde \mu \frac{\xi^0_{,1}}{B}- \tilde q \left(\xi^0_{,0} + \xi^0 \frac{\dot B}{B} + \xi^1 \frac{B^\prime}{B} + \xi^1_{,1}\right) + \epsilon B \xi^1_{,0} = 0,
\label{eno2}
\end{equation}
\begin{equation}
\xi^0 \dot \epsilon + \xi^1 \epsilon^\prime - 2 \tilde q \frac{\xi^0_{,1}}{B} + 2\epsilon \left( \xi^0 \frac{\dot B}{B} + \xi^1 \frac{B^\prime}{B} + \xi^1_{,1}\right) = 0.
\label{eno3}
\end{equation}

We shall consider two separated subcases, namely:
\begin{itemize}
\item Dissipation in purely diffusion approximation ($q\neq0, \epsilon=0, \tilde q=q$).
\item Dissipation in the streaming out limit ($\epsilon\neq0, q=0, \tilde q=\epsilon$).
\end{itemize}

\subsection{Diffusion approximation}

\noindent From (\ref{13}) and  (\ref{HDO}),
we can write:
\begin{equation}
8\pi  q B=\frac{\dot
K}{1+K}\frac{R^\prime}{R}.\label{e1}
\end{equation}

Combining   (\ref{HDO}), (\ref{difn1}) and (\ref{e1}) produces
\begin{equation}
8 \pi g(r) =R R^{\prime 2} \frac{\dot K}{(1+K)^{3/2}}.
\label{kpg}
\end{equation}

Next, from  (\ref{eno3})  we obtain $\xi^0=F(t)$, as in LTB, and  (\ref{eno1}) and (\ref{eno2}) take the form
\begin{equation}
F(t) \dot{ \mu} + \xi^1 \mu^\prime+2\mu \dot F(t) - 2 q B \xi^1_{,0} =0,
\label{eno1c0}
\end{equation}
\begin{equation}
F(t) \dot {q} + \xi^1 {q}^\prime + q \left(\dot F(t) +  F(t) \frac{\dot B}{B} + \xi^1 \frac{B^\prime}{B} + \xi^1_{,1}\right) = 0.
\label{eno2c0}
\end{equation}
 This last equation  can be rewritten in the form
\begin{equation}
\xi^1 \left[\ln{\left(q B \xi^1\right)}\right]^\prime + F(t) \left[\ln\left(q B F(t)\right)\right]^{\dot{}}=0.
\label{eno2cp}
\end{equation}
\noindent Multiplying (\ref{eno2cp})
by  $qB$, it becomes:
\begin{equation}
(qB\xi^1)^\prime +(qBF\dot )=0,\label{e2}
\end{equation}
\noindent a partial solution of which may be writen as

\begin{eqnarray}
q B \xi^1 &=&-\dot \psi(t,r) \label{int1},\\
q B F(t) &=& \psi ^\prime (t,r).\label{e3}
\end{eqnarray}

\noindent From  (\ref{e1})  and  (\ref{e3}) we have:
\begin{equation}
\psi ^\prime (t,r)=\frac {F(t)}{8\pi}\frac{\dot
K}{1+K}\frac{R^\prime}{R}.\label{intnue}
\end{equation}

Then, using (\ref{difn1}) in (\ref{e3}) with (\ref{intnue}), and taking (\ref{HDO}) into account, we get
\begin{equation}
8 \pi g(r) =R R^{\prime 2} \frac{\dot K}{(1+K)^{3/2}},
\label{kpgn}
\end{equation}
which is exactly (\ref{kpg}). 

More generally, it can be shown that integrability conditions for (\ref{eno1c0}) and (\ref{e2}) can always be satisfied. In the particular case   $F(t)=1$, we obtain

\begin{equation}\nonumber \xi^1 = - \frac{(\ln qB)^{\ddot{}}  + \frac12\left( \frac{\mu^\prime}{qB}\right)^{\dot{}}}{(\ln qB)^{\dot{} \prime}  + \frac12\left( \frac{\mu^\prime}{qB}\right)^{\prime}}.\label{nuevaJC}
\end{equation}

In other words the existence of a proper matter collineation is consistent with the general form of GLTB in the diffusion limit.

We can  integrate  (\ref{kpg}) to obtain
\begin{equation}
\frac{1}{(1+K)^{1/2}}=-4 \pi g(r) \int{\frac{dt}{R R^{\prime 2}}}+c(r),
\label{ink}
\end{equation}
where $c(r)$ is an arbitrary function.

As a specific example, let us choose for $R(t,r)$ the form for the parabolic subclass of LTB  given by (\ref{RT}),  then we obtain
\begin{widetext}
\begin{equation}
\frac{1}{(1+K)^{1/2}}=-\frac{4 \pi g(r)}{f(r) f^{\prime}(r)\left[f^{\prime}(r)(T(r)-t)+\frac{2}{3}f(r)T^{\prime}(r)\right]}+c(r),
\label{1mK}
\end{equation}
and introducing this last expression into (\ref{difn1})
we have
\begin{equation}
q=\frac{g(r)}{f^2(r)(T(r)-t)^{2/3}}\left\{-\frac{4 \pi g(r)}{f(r)f^{\prime}(r)}+c(r)\left[f^{\prime}(r)(T(r)-t)+\frac{2}{3}f(r)T^{\prime}(r)\right]\right\}^{-2}.
\label{qeo}
\end{equation}
\end{widetext}
where we have used (\ref{HDO}).

So far the  GLTB thus obtained, is defined up to four functions of $r$, which by the invariance with respect to the transformation of the radial coordinate reduces to three. Two of them can be taken exactly as the  corresponding in the LTB ``seed'' model, and the remaining one is obtained via equations (\ref{e2})--(\ref{e3}), from the assumption that $\xi^1$ has the same radial dependence as in the LTB case. Observe that due to the junction condition (\ref{j3}) we have to impose $g(r)\stackrel{\Sigma}{=}0$.
Finally, the temperature profile can be easily obtained from  (\ref{ISE'}) (or (\ref{ISE'bis})) and (\ref{qeo}).
\subsection{Streaming out approximation}
In the case $q=0$ and $\epsilon \not=0$,  we obtain from Bianchi identities (\ref{an}) and (\ref{bn})
\begin{eqnarray}
\dot \mu+\mu\left(\frac{\dot B}{B}+2\frac{\dot R}{R}\right)=0,\label{1an'}
\end{eqnarray}
producing
\begin{equation}
 \mu=\frac{j(r)}{BR^2},\label{1nltb}
\end{equation}
exactly as in the LTB case.
However, using (\ref{HDO})--(\ref{Kq1}) we get in this case
\begin{equation}
 \mu=\frac{3 j(r)\left[\int 4\pi\epsilon Rdt+
(1+\kappa(r))^{1/2}\right]}{(R^{3})^\prime},\label{n1ltb}
\end{equation}
where $j(r)$ is a function of integration.

Next, in this approximation (\ref{eno1})--(\ref{eno3}) take the form
\begin{equation}
\xi^0 (\mu+\epsilon)^{\dot{}} + \xi^1(\mu+\epsilon)^\prime+2(\mu+\epsilon) \xi^0_{,0}- 2 \epsilon B \xi^1_{,0} =0,
\label{eno11}
\end{equation}
\begin{eqnarray}
\xi^0 \dot {\epsilon} + \xi^1 {\epsilon}^\prime -(\mu+\epsilon) \frac{\xi^0_{,1}}{B} \nonumber \\+ \epsilon \left(\xi^0_{,0} + \xi^0 \frac{\dot B}{B} + \xi^1 \frac{B^\prime}{B} + \xi^1_{,1}\right) - \epsilon B \xi^1_{,0} = 0,
\label{eno22}
\end{eqnarray}
\begin{equation}
\xi^0 \dot \epsilon + \xi^1 \epsilon^\prime - 2 \epsilon \frac{\xi^0_{,1}}{B} + 2\epsilon \left( \xi^0 \frac{\dot B}{B} + \xi^1 \frac{B^\prime}{B} + \xi^1_{,1}\right) = 0.
\label{eno33}
\end{equation}
From (\ref{eno11})--(\ref{eno33}) we find
\begin{equation}
\xi^0 \dot\mu + \xi^1 \mu^\prime + 2 \mu \xi^0_{,0} + 2 \mu \frac{\xi^0_{,1}}{B}=0,
\label{a}
\end{equation}
 producing
\begin{equation}
\xi^1=-\frac{\mu}{\mu^\prime}\left(2\dot F(r,t) + F(r,t)\frac{\dot \mu}{\mu} + \frac{2 F^\prime(r,t)}{B}\right),
\label{bb}
\end{equation}
with $\xi^0=F(r,t)$.

Following   the requirement  of ``maximal'' similarity between a LTB and its  corresponding GLTB,  let us assume  $\xi^0=F(t)$, in which case   (\ref{bb}), becomes  (\ref{xi1}).

Under that condition, we have from (\ref{eno11})--(\ref{eno33})
\begin{equation}
F(t) (\mu+\epsilon)^{\dot{}} + \xi^1(\mu+\epsilon)^\prime+2(\mu+\epsilon) \dot{F}(t)- 2 \epsilon B \xi^1_{,0} =0,
\label{eno111}
\end{equation}
\begin{eqnarray}
F(t) \dot {\epsilon} + \xi^1 { \epsilon}^\prime+ \epsilon \left(\dot{F}(t) + F(t) \frac{\dot B}{B} + \xi^1 \frac{B^\prime}{B} + \xi^1_{,1}\right)\nonumber \\ - \epsilon B \xi^1_{,0} = 0,
\label{eno222}
\end{eqnarray}
\begin{equation}
F(t) \dot \epsilon + \xi^1 \epsilon^\prime  + 2\epsilon \left( F(t) \frac{\dot B}{B} + \xi^1 \frac{B^\prime}{B} + \xi^1_{,1}\right) = 0.
\label{eno333}
\end{equation}
This last equation can be written as:
\begin{equation}
F(t)  \left[\ln\right(\epsilon B^2\left)\right]^{\dot{}}+ \xi^1 \left[\ln\right(\epsilon(B\xi^1)^2\left)\right]^\prime=0,
\label{33}
\end{equation}
or in the form
\begin{equation}
F(t)(\epsilon B^2)^{\dot{}}+\frac{1}{\xi^1}(\epsilon (B \xi^1)^2)^\prime=0.
\label{4as}
\end{equation}
Also, in this case, we can write (\ref{bn}) in the form
\begin{equation}
 \left[\ln\right(\epsilon (BR)^2\left)\right]^{\dot{}}+ \frac{1}{B} \left[\ln\right(\epsilon R^2\left)\right]^\prime=0,
\label{bnqo}
\end{equation}
or,
\begin{equation}
\left[\epsilon (BR)^2\right]^{\dot{}}+B(\epsilon R^2)^\prime=0.
\label{5as}
\end{equation}
The method to obtain GLTB from any given LTB (in the streaming out approximation) may now be sketched as follows:
\begin{itemize}
\item Take any specific ``seed'' LTB with a given vector field $\xi$ defining a proper matter collineation (since $F(t)$ is arbitrary we put it equal to one).
\item Replace the term $(\epsilon B^2)^{\dot{}}$ in (\ref{4as}) by its expression obtained from (\ref{5as}).
\item Assume that the radial dependence of $\xi^1$ is the same as the corresponding to the LTB ``seed'' solution.
\item From the form of $R$ corresponding to the ``seed'' LTB, integrate (\ref{4as}) with respect to $r$ to obtain $\epsilon$.  Any remaining arbitrary function of $t$  may be found from junction conditions.
\end{itemize}
\section{CONCLUSIONS}
We have presented different alternatives for obtaining spherically symmetric, geodesic, dissipative dust  solutions to Einstein equations (GLTB). All these alternatives are oriented to produce solutions with specific similarities to the non--dissipative case (LTB). For doing so we have first carried out a study on LTB spacetimes based, on the one hand on some symmetry properties of LTB, and on the other on their characterization in terms of the structure scalars.

The generalizations based on structure scalars produce a dissipative model for each ``seed'' LTB, such that dissipative variables are arbitrary only up to a single function of $r$.

However, the generalizations based on symmetry properties have more degrees of freedom, which may be specified depending on the specific problem under consideration.

For the purely diffusion case the temperature profile may be obtained from the transport equation considered in section VI.

At the present time we foresee two interesting  applications of GLTB's: One the  testing  of any result obtained from any specific LTB against the presence of small but nonvanishing dissipative fluxes, such dissipative perturbations being represented  by  exact solutions to the Einstein equations; and two, the study of relaxational effects on different aspects of collapse via the transport equation.

Finally, as already stressed in the Introduction, let us recall that our intention here is not to produce specific solutions  with a distinct physical interpretation, but rather to pave the way for finding such solutions.

\begin{acknowledgments}
LH wishes to thank Fundaci\'on Empresas Polar for financial support and Universitat de les  Illes Balears  and Departamento de F\'isica Fundamental at Universidad de Salamanca, for financial support and hospitality. ADP  acknowledges hospitality of the
Departament de F\'isica at the  Universitat de les  Illes Balears and Departamento de F\'isica Fundamental at Universidad de Salamanca.

\end{acknowledgments}


\begin{thebibliography}{99}
\bibitem{1} G. Lema\^{\i}tre {\it Ann. Soc. Sci. Bruxelles} {\bf A 53}, 51 (1933).

\bibitem{2} R. C. Tolman {\it Proc. Natl. Acad Sci} {\bf 20}, 169  (1934).

\bibitem{3} H. Bondi {\it  Mon. Not. R. Astron. Soc.} {\bf 107}, 410 (1947).

\bibitem{4} A. Krasinski  {\it Inhomogeneous Cosmological Models},(Cambridge University Press, Cambridge) (1998).
\bibitem{5} J. Plebanski and A. Krasinski  {\it An Introduction to General Relativity  and Gravitation},(Cambridge University Press, Cambridge) (2006).
\bibitem{ns} R. Sussman, in {\it Gravitation and Cosmology: Proceedings
of the Third International Meeting on Gravitation and
Cosmology}, edited by A. Herrera--Aguilar, F. S. Guzman
Murillo, U. Nucamendi Gomez, and I. Quiros, AIP Conf.
Proc. 1083 (AIP, New York, 2008), pp. 228-235.
\bibitem{sn} R. Sussman and G. Izquierdo  {\it  arXiv:1004.0773}.

\bibitem{8} R. Maartens and  D. R. Matravers  {\it arXiv: gr-qc/9804023v1}.
\bibitem{7} D. R. Matravers and N. P. Humphreys {\it Gen. Rel. Grav.} {\bf 33}, 531 (2001).
\bibitem{6} C. Hellaby and A. Krasinski {\it Phys. Rev. D} {\bf 73}, 023518 (2006).
\bibitem{9} D. M. Eardley and L. Smarr  {\it Phys. Rev. D} {\bf 19}, 2239 (1979).
\bibitem{10}  B. Waugh and K. Lake   {\it Phys. Rev. D} {\bf 40}, 2137 (1989).

\bibitem{11} P. S. Joshi and I. H. Dwivedi  {\it Phys. Rev. D} {\bf 47}, 5357 (1993).
\bibitem{12} P. S. Joshi and T. P. Singh {\it Phys. Rev. D} {\bf 51}, 6778 (1995).
\bibitem{J} P. S. Joshi, N. Dadhich and R. Maartens {\it Phys. Rev. D} {\bf 65}, 101501(R) (2002).
\bibitem{m1} J. Mimoso, M. Le Delliou and F. Mena  {\it Phys. Rev. D} {\bf 81}, 123514 (2010).
\bibitem{m2}  M. Le Delliou, F. Mena and J. Mimoso,  {\it arXiv: 0911.0241}.
\bibitem{13} C. Vaz, L. Witten and T. P. Singh {\it Phys. Rev. D} {\bf 63}, 104020 (2001).
\bibitem{14} M. Bojowald, T. Harada and R. Tibrewala  {\it Phys. Rev. D} {\bf 78}, 064057 (2008).
\bibitem{Coley1} A. A. Coley and N. Pelavas {\it Phys. Rev. D} {\bf 74}, 087301 (2006).
\bibitem{Coley2} A. A. Coley and N. Pelavas {\it Phys. Rev. D} {\bf 75}, 043506 (2007).

\bibitem{17p} M. N. Celerier {\it New Advances in Physics} {\bf 1}, 29 (2007).
\bibitem{15} S. Viaggiu {\it arXiv:0907.0600v1}.
\bibitem{7''} R. Sussman {\it arXiv:0807.1145v2}.
\bibitem{cel} M. N. Celerier {\it arXiv:0911.2597v}
\bibitem{7N} R. Sussman {\it arXiv:1001.0904v1}.
\bibitem{Hs} L. Herrera and N. O. Santos  {\it Phys. Rev. D} {\bf 70}
084004 (2004).
\bibitem{Hetal} L. Herrera, A.  Di Prisco, J. Mart\'\i n, J. Ospino, N. O. Santos
and O. Troconis {\it Phys. Rev. D} {\bf 69} 084026 (2004).
\bibitem{Mitra} A. Mitra {\it Phys. Rev. D} {\bf 74} 024010 (2006).
\bibitem{K} D. Kazanas  and D. Schramm {\it Sources of Gravitational
Radiation}, L. Smarr ed.,
(Cambridge University Press, Cambridge) (1979).
\bibitem{3n} W. D. Arnett {\it Astrophys. J.}  {\bf 218} 815 (1977).
\bibitem{4n} D. Kazanas {\it Astrophys. J.} {\bf 222} L109 (1978).
\bibitem{5n} J. Lattimer {\it Nucl. Phys.} {\bf A478} 199 (1988).
\bibitem{19} C. Kolassis, N.  O. Santos and D. Tsoubelis {\it Astrophys. J.} {\bf 327}, 755 (1988).
\bibitem{20} M. Govender, S. Maharaj and R. Maartens {\it Class. Quantum Grav.} {\bf 15}, 323 (1998).
\bibitem{22} L. Herrera, G. Le Denmat and N. O. Santos {\it Int. J. Mod. Phys. D}
{\bf 13} 583 (2004).
\bibitem{18} S. Maharaj, M. Govender {\it Int. J. Mod. Phys. D} {\bf 14}, 667 (2005).
\bibitem{17} S. Thirukkanesh and S. Maharaj {\it J. Math. Phys.} {\bf 50}, 022502 (2009).
\bibitem{21} N. Naidu, M. Govender and K. Govinder {\it Int. J. Mod. Phys. D} {\bf 15}, 1053 (2006).
\bibitem{24} S. Rajah and S. Maharaj {\it J. Math. Phys.} {\bf 49}, 012501 (2008).
\bibitem{26} G. Darmois {\it M\'emorial des Sciences Math\'ematiques}
(Gauthier-Villars, Paris, 1927) Fasc. 25.

\bibitem{Herreraanis}L. Herrera and N. O. Santos {\it Phys. Rep.} {\bf 286}, 53 (1997).

\bibitem{Misner} C. Misner and D. Sharp, {\it Phys. Rev.} {\bf 136}, B571 (1964).
\bibitem{Cahill} M. Cahill and G. McVittie,  {\it J. Math. Phys.} {\bf 11}, 1382 (1970).
\bibitem{chan1}R. Chan {\it Mon. Not. R. Astron. Soc.} {\bf 316}, 588 (2000).
\bibitem{Santos}N. O. Santos {\it Mon. Not. R. Astron. Soc.} {\bf 216}, 403 (1985).
\bibitem{Bonnor}W. B. Bonnor, A. Oliveira and N. O. Santos, {\it Phys. Rep.} {\bf 181}, 269 (1989).
\bibitem{ellis1} G. F. R. Ellis, {\ Relativistic Cosmology} in: Proceedings of the International School of Physics `` Enrico Fermi'', Course 47: General Relativity and Cosmology. Ed. R. K. Sachs (Academic Press, New York and London) (1971).
\bibitem{ellis2} G. F. R. Ellis, {\it Gen. Rel. Grav.} {\bf  41}, 581 (2009).
\bibitem{she} L. Herrera, A. Di Prisco and J. Ospino {\it Gen.Rel. Grav.}{\bf 42}, 1585 (2010).
\bibitem{H1} L. Herrera, N. O. Santos and A. Wang {\it Phys. Rev. D} {\bf 78}, 084026 (2008).

\bibitem{sp} L. Herrera, J. Ospino, A. Di Prisco, E. Fuenmayor and O. Troconis, {\it Phys. Rev. D} {\bf 79}, 064025 (2009).
\bibitem{bel1} L. Bel, {\it Ann. Inst. H
Poincar\'e}  {\bf 17}, 37 (1961).
\bibitem{parrado} A. Garc\'ia--Parrado G\'omez Lobo, {\it arXiv:0707.1475v2}.
\bibitem{s1} R. A. Sussman, {\it Phys. Rev. D } {\bf 79}, 025009 (2009).

\bibitem{Carot94} J. Carot, J. da Costa and E.G.L.R. Vaz, {\it J. Math. Phys.} {\bf 35}, 4832 (1994).

\bibitem{Muller67}  I M\"{u}ller {\it Z. Physik} {\bf 198}, 329 (1967)
\bibitem{IsSt76} W  Israel {\it Ann. Phys.} (NY) {\bf 100}, 310 (1976).
\bibitem{I} W. Israel and J. Stewart {\it Phys. Lett. A} {\bf 58}.
213  (1976).
\bibitem{II} W. Israel and J. Stewart {\it Ann. Phys.} (NY) {\bf 118}, 341 (1979).

\bibitem{6D} D Joseph  and L Preziosi {\it Rev. Mod. Phys.}
{\bf 61}, 41 (1989)

\bibitem{7D} D. Jou, J Casas-V\'azquez and G Lebon {\it Rep. Prog. Phys.}
{\bf 51}, 1105 (1988)

\bibitem{8d} R Maartens {\it astro-ph/9609119}

\bibitem{8'} L Herrera and D Pav\'on
{\it Physica A}, {\bf 307}, 121 (2002)

\bibitem{9H} W Hiscock  and L Lindblom {\it Ann. Phys.} (NY)
{\bf 151}, 466 (1983)

\bibitem{10E} C Eckart {\it Phys. Rev.} {\bf 58}, 919 (1940)

\bibitem{11L} L Landau and E Lifshitz, {\it Fluid Mechanics}
(Pergamon Press, London) (1959)

\bibitem{14Di} D Pav\'on, D Jou  and J Casas-V\'azquez {\it Ann. Inst. H
Poincar\'e} {\bf A36}, 79 (1982)

\bibitem{15d} B Carter {\it Journ\'ees Relativistes}, ed. M Cahen, R
Debever and J Geheniau, (Universit\'e Libre de Bruxelles) (1976)

\bibitem{18D} C Cattaneo {\it Atti Semin. Mat. Fis. Univ. Modena}
{\bf 3}, 3 (1948)
\bibitem{pavon}J. Triginer and D. Pavon  {\it Class. Quantum Grav.} {\bf 12}, 689, (1995).
\end{thebibliography}
\end{document}